\documentclass[10pt,twocolumn]{article}

\usepackage[T1]{fontenc}
\usepackage[utf8]{inputenc}
\usepackage{lmodern}
\usepackage[a4paper,margin=1.8cm,columnsep=0.7cm]{geometry}
\usepackage{graphicx}
\usepackage{booktabs}
\usepackage{array}
\usepackage{xcolor}
\usepackage{enumitem}
\usepackage{caption}
\usepackage[numbers,round,sort&compress]{natbib}
\usepackage{tikz}
\usetikzlibrary{positioning,arrows.meta}
\usepackage[hidelinks]{hyperref}
\usepackage{xurl}
\usepackage{microtype}

\newcommand{\kaggleurl}{https://www.kaggle.com/competitions/unifesp-fatty-liver}

\title{\textbf{The UNIFESP Chest CT Fatty Liver Dataset}}

\author{%
Augusto Sarquis Serpa$^{1,2,3}$, Mariana Ramos Scolari$^{1,4}$,\\
Eduardo Moreno Judice de Mattos Farina$^{1,5}$, Eduardo Caminha Nunes$^{1,4}$,\\
Felipe Matsuoka$^{6}$, Felipe Campos Kitamura$^{1,7}$, Nitamar Abdala$^{1}$\\[0.6em]
\small $^{1}$Department of Diagnostic Imaging, Escola Paulista de Medicina,\\
\small Universidade Federal de S\~ao Paulo (UNIFESP), S\~ao Paulo, Brazil\\
\small $^{2}$Hospital Israelita Albert Einstein, S\~ao Paulo, Brazil\\
\small $^{3}$JC Imagens, Cuiab\'a, Brazil\\
\small $^{4}$Rede D'Or S\~ao Luiz, S\~ao Paulo, Brazil\\
\small $^{5}$University of Texas Southwestern Medical Center, Dallas, Tex, USA\\
\small $^{6}$Dasa (Diagn\'osticos da Am\'erica S.A.), S\~ao Paulo, Brazil\\
\small $^{7}$Eden, Mexico City, Mexico\\[0.3em]
\footnotesize The work was performed at UNIFESP (affiliation 1). Affiliations 2--5 are the
current affiliations\\
\footnotesize of authors who were at UNIFESP at the time of the work.\\[0.3em]
\small Corresponding author: Augusto Sarquis Serpa (\texttt{augusto.esd5@gmail.com})
}
\date{}

\begin{document}
\maketitle

\begin{abstract}
\noindent
Hepatic steatosis is common, usually asymptomatic, and frequently visible on CT examinations
performed for other reasons, including chest CT, in which the upper abdomen is partly included.
We describe the UNIFESP Chest CT Fatty Liver Dataset, a public collection of 285 unenhanced CT
studies (344 series, 99\,883 images) from 264 patients examined at a single academic center in
Brazil: 152 patients with chest CT and 112 patients with abdominal CT. Each patient has a binary
label for the presence of hepatic steatosis, assigned by a radiology resident who compared the
attenuation of three regions of interest in the liver with one in the spleen. Images are provided
in de-identified DICOM format. The dataset was used in the UNIFESP Chest CT Fatty Liver
Competition, hosted on Kaggle in 2022, with 226 patients for training and 38 patients with chest
CT for testing. In a technical validation, the criterion of liver attenuation lower than spleen
attenuation, measured automatically, agreed with the labels in 206 of the 226 training patients
(91\%). The dataset is available at \url{\kaggleurl}.
\end{abstract}


\section*{Summary}

The UNIFESP Chest CT Fatty Liver Dataset contains 285 unenhanced chest and abdominal CT studies
from 264 patients with patient-level labels for hepatic steatosis and is available at
\url{\kaggleurl}.

\section*{Key Points}

\begin{itemize}[leftmargin=1.5em]
\item The UNIFESP Chest CT Fatty Liver Dataset is a publicly available collection of 285
unenhanced CT studies (99\,883 images) from 264 patients at a single academic center in Brazil,
with 152 patients who underwent chest CT and 112 who underwent abdominal CT.
\item Each patient has a binary label for hepatic steatosis, assigned by a radiology resident on
the basis of attenuation measurements in the liver and the spleen; 147 of the 226 patients (65\%)
in the training set are labeled as having steatosis.
\item The dataset was used for the UNIFESP Chest CT Fatty Liver Competition on Kaggle, in which
models were tested only on chest CT, and it supports research on the opportunistic detection of
steatosis in examinations not centered on the liver.
\end{itemize}

\section*{Introduction}

Steatotic liver disease is the most common chronic liver disease, with an estimated global
prevalence of about 30\% for its main form, which is now called metabolic
dysfunction--associated steatotic liver disease \citep{younossi2023,rinella2023delphi}. Most
affected individuals have no symptoms, yet a proportion progress to steatohepatitis, fibrosis,
cirrhosis, and hepatocellular carcinoma, and the condition is closely linked to cardiometabolic
disease \citep{rinella2023guidance}. Because the disease is silent, imaging examinations
performed for unrelated reasons are an opportunity to identify it.

At unenhanced CT, the liver of healthy adults is slightly more attenuating than the spleen
\citep{piekarski1980}, and the attenuation of the liver decreases as its fat content increases.
Steatosis is therefore assessed by the absolute attenuation of the liver, by the difference
between liver and spleen attenuation, or by their ratio
\citep{park2006,kodama2007,starekova2021}. Liver attenuation is highly specific for moderate to
severe steatosis but has low sensitivity \citep{pickhardt2012}, and CT is inaccurate for mild
steatosis \citep{lee2014}.

Chest CT is one of the most frequently performed CT examinations, and it almost always includes
the upper part of the liver and of the spleen. Liver fat can be measured on both chest and abdominal CT scans, although with lower precision
on chest scans \citep{speliotes2008}, and the liver-minus-spleen attenuation difference performs similarly
on chest and abdominal protocols \citep{li2025}. Steatosis has accordingly been described in
lung cancer screening programs \citep{chen2017} and in patients who underwent chest CT for
COVID-19 \citep{medeiros2020}. However, the reader of a chest CT is focused on the thorax, and a
mild decrease in liver attenuation is easily overlooked unless it is actively measured; even on
unenhanced abdominal CT, steatosis that is measurable on the images is often absent from the
report \citep{yardeni2026}. Automated measurement of liver attenuation is thus a natural
application of image analysis and machine learning, and several automated methods have been
proposed for abdominal and chest CT \citep{graffy2019,jirapatnakul2020,modanwal2022,weiss2025}. These methods were developed
with institutional or controlled-access data. Public image datasets with steatosis labels exist
for ultrasound \citep{byra2018}, but, to our knowledge, no openly downloadable CT dataset
provides radiologist-assigned labels for hepatic steatosis on chest CT.

In other areas of radiology, public datasets created for image analysis competitions have
supported the development of machine learning models \citep{colak2021,rudie2024}. The purpose of this work
was to create and make publicly available a dataset of unenhanced chest and abdominal CT
examinations with labels for hepatic steatosis, to release it through a machine learning
competition \citep{kagglecompetition}, and to describe its composition and the consistency of
its labels.

\section*{Materials and Methods}

\subsection*{Study Design and Data Collection}

This retrospective study was performed at a single academic center, the Department of
Diagnostic Imaging of the Escola Paulista de Medicina, Universidade Federal de S\~ao Paulo
(UNIFESP), S\~ao Paulo, Brazil, and was approved by the institutional research ethics committee,
with a waiver of informed consent. A convenience sample of chest and abdominal CT examinations
was retrieved from the institutional picture archiving and communication system. As the first
step of the process, before review and annotation, the Digital Imaging and Communications in
Medicine (DICOM) files were de-identified with the RSNA Anonymizer (Radiological Society of North
America). The examination dates, the age and sex of the patients, and the manufacturer and model
of the scanners were not retained in the de-identified files and were not recorded separately.

\subsection*{Annotation}

All examinations were reviewed by one second-year radiology resident. For each examination, the
resident drew three regions of interest (ROIs) in the liver parenchyma and one ROI in the spleen
and compared the attenuation of the two organs. Steatosis was recorded as present when the
attenuation of the liver was lower than that of the spleen, regardless of the magnitude of the
difference; no minimum difference or absolute attenuation threshold was applied. Only the binary
result (steatosis present or absent) was recorded, and one label was assigned per patient. The
size and position of the ROIs, the series used when more than one section thickness was
available, and the way in which patients with more than one study were handled were not
recorded. Examinations with very large or diffuse lesions of the liver parenchyma that prevented
a reliable measurement were excluded.

\subsection*{Training and Test Sets and Competition}

The dataset was prepared for the UNIFESP Chest CT Fatty Liver Competition, a community
prediction competition hosted on Kaggle from April 6 to December 31, 2022
\citep{kagglecompetition}. Patients were divided into a training set, for which the labels were
released, and a test set, for which the labels were withheld. All patients with abdominal CT
were assigned to the training set, so that models could learn from examinations in which the
whole liver is depicted, and the test set was composed only of patients with chest CT, the
examination of interest. Participants submitted, for each patient in the test set, a predicted
probability of steatosis. The test set was divided by the platform into public and private
portions of approximately equal size for the leaderboards, and submissions were ranked by the
binary cross-entropy (log loss),
\[
\mathrm{LL} = -\frac{1}{N}\sum_{i=1}^{N}\left[y_i \log p_i + (1-y_i)\log(1-p_i)\right],
\]
where $N$ is the number of patients in the test set, $y_i$ is the label of patient $i$, and
$p_i$ is the predicted probability of steatosis. The competition rules asked participants not to
annotate the test set themselves.

\subsection*{Data Organization and Availability}

CT images are provided in DICOM format. Files are organized in a directory tree with one folder
per case, named with a sequential case number, which contains one folder per study and, within
it, one folder per series. The Study Description attribute of each file holds the type of
examination (``Chest'' or ``Abdomen''). Two comma-separated value files accompany the images:
\texttt{train.csv} lists the case number (\texttt{Id}) and the label (\texttt{ground\_truth}, 1
for steatosis and 0 for no steatosis) of each case in the training set, and
\texttt{sample\_submission.csv} lists the case numbers of the test set. Data are available at
\url{\kaggleurl}; access requires a Kaggle account, and use of the data is subject to the rules
of the competition. Predictions for the test set can be scored through the competition page.
To verify the integrity of the released files and to tabulate the acquisition and
reconstruction parameters of each series, the header of every DICOM file was read with pydicom
\citep{pydicom}.

\subsection*{Technical Validation}

To assess the consistency of the labels, liver and spleen attenuation were measured
automatically in every study. These measurements were made for this report, after the
competition, on the files as released, and were not used to assign the labels. For each study,
we selected the series with the largest section thickness, converted the images to Hounsfield
units, and resampled the volume to 3-mm isotropic voxels (every $n$th section was taken to
approximate 3-mm spacing, followed by linear interpolation). The liver, spleen, and aorta were
segmented with the 3-mm model of TotalSegmentator \citep{wasserthal2023}, an nnU-Net
\citep{isensee2021}, without manual correction. Each mask was eroded in the axial plane (6 mm for
the liver and spleen, 3 mm for the aorta) to exclude the organ boundary, and the mean
attenuation of the remaining voxels was recorded. Because the whole organ is averaged, the liver
value includes intrahepatic vessels and any focal lesion. The attenuation of the aorta was used
to check for the presence of intravenous contrast material. An organ was considered truncated
when its mask reached the most caudal section of the volume.

The automated measurements were compared with the labels in the training set. Each study was
given the label of its patient, and a study was considered positive by the automated criterion
when the mean attenuation of the liver was lower than that of the spleen. For the patient-level
comparison, a patient was considered positive when this criterion was met in at least one study.
Agreement was summarized as the proportion of concordant cases and with the Cohen $\kappa$, and
the discrimination of the liver-minus-spleen difference for the label was summarized with the
area under the receiver operating characteristic curve. Continuous variables are reported as
medians with IQRs or ranges. No hypothesis tests were performed.

\section*{Results}

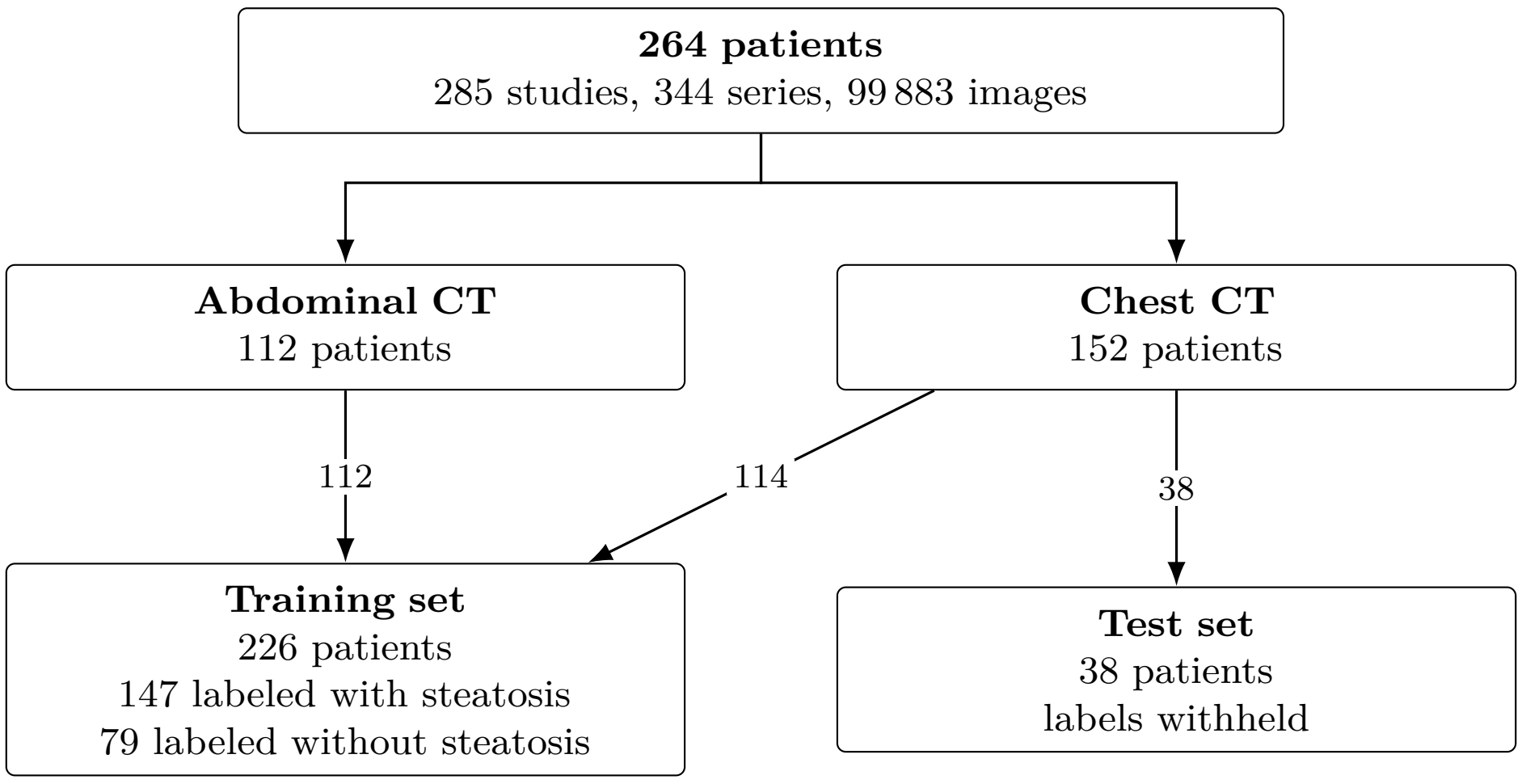
\begin{figure*}[t]
\centering
\begin{tikzpicture}[
  box/.style={draw, rounded corners=2pt, align=center, font=\small, inner sep=5pt, text width=#1},
  box/.default=5.2cm,
  arr/.style={-{Latex[length=2mm]}, semithick},
  lab/.style={font=\footnotesize, fill=white, inner sep=1.5pt}]
\node[box=8.2cm] (all) at (0,0) {\textbf{264 patients}\\ 285 studies, 344 series, 99\,883 images};
\node[box] (abd) at (-3.4,-2.1) {\textbf{Abdominal CT}\\ 112 patients};
\node[box] (chest) at (3.4,-2.1) {\textbf{Chest CT}\\ 152 patients};
\node[box] (train) at (-3.4,-4.9) {\textbf{Training set}\\ 226 patients\\ 147 labeled with steatosis\\ 79 labeled without steatosis};
\node[box] (test) at (3.4,-4.9) {\textbf{Test set}\\ 38 patients\\ labels withheld};
\draw[arr] (all.south) -- ++(0,-4mm) -| (abd.north);
\draw[arr] (all.south) -- ++(0,-4mm) -| (chest.north);
\draw[arr] (abd.south) -- node[lab] {112} (abd.south |- train.north);
\draw[arr] (chest.south) -- node[lab] {38} (chest.south |- test.north);
\draw[arr] (chest.south west) ++(8mm,0) -- node[lab] {114} ([xshift=-8mm]train.north east);
\end{tikzpicture}
\caption{Composition of the dataset and division into the training and test sets used in the
competition. Numbers on the arrows are numbers of patients. The 152 patients in the chest CT
group include four who also have an abdominal CT study.}
\label{fig:flow}
\end{figure*}

\begin{table*}[t]
\centering
\small
\caption{Composition of the Training and Test Sets}
\label{tab:composition}
\begin{tabular}{@{}lrrrr@{}}
\toprule
 & \multicolumn{2}{c}{Training set} & Test set & \\
\cmidrule(lr){2-3}\cmidrule(lr){4-4}
Characteristic & Chest CT & Abdominal CT & Chest CT & Total \\
\midrule
Patients & 114 & 112 & 38 & 264 \\
Patients with more than one study & 7 & 8 & 3 & 18 \\
Studies & 122 & 121 & 42 & 285 \\
Series & 128 & 173 & 43 & 344 \\
Images & 40\,772 & 44\,692 & 14\,419 & 99\,883 \\
\addlinespace
Patients labeled with steatosis & 64 (56) & 83 (74) & Withheld & \ldots \\
Patients labeled without steatosis & 50 (44) & 29 (26) & Withheld & \ldots \\
\bottomrule
\multicolumn{5}{@{}p{0.92\linewidth}@{}}{\footnotesize Note.---Data are numbers of patients,
studies, series, or images, with percentages in parentheses. Patients are grouped by type of
examination; the chest CT group of the training set includes four patients who also have an
abdominal CT study (seven studies and 10 series in these four patients).}
\end{tabular}
\end{table*}

\subsection*{Dataset Composition}

The dataset is composed of 285 CT studies in 264 patients, with a total of 344 image series and
99\,883 images (52.5 GB uncompressed). Of the 264 patients, 152 underwent chest CT (four of whom
also have an abdominal CT study) and 112 underwent abdominal CT only. Eighteen patients have more
than one study (15 patients have two and three patients have three), and 58 studies include more
than one series, which correspond to reconstructions of the same acquisition with different
section thicknesses.

\subsection*{Steatosis Labels}

A label was assigned to all 264 patients. In the 226 patients whose labels are public, 147
(65\%) are labeled as having steatosis and 79 (35\%) as not having steatosis. Steatosis was
recorded in 83 of the 112 patients with abdominal CT (74\%) and in 64 of the 114 patients with
chest CT whose labels are public (56\%). The number of examinations excluded during annotation
was not recorded.

\begin{table*}[t]
\centering
\small
\caption{Technical Characteristics of the Image Series}
\label{tab:technical}
\begin{tabular}{@{}lrr@{}}
\toprule
Characteristic & Chest CT series ($n$ = 165) & Abdominal CT series ($n$ = 179) \\
\midrule
Section thickness & & \\
\quad 1 mm & 162 & 56 \\
\quad 2 mm & 0 & 2 \\
\quad 3 mm & 0 & 114 \\
\quad 5 mm & 3 & 7 \\
Tube voltage & & \\
\quad 120 kVp & 164 & 179 \\
\quad 140 kVp & 1 & 0 \\
Images per series & 330 (60--435) & 168 (60--670) \\
Craniocaudal coverage (mm) & 329 (79--434) & 474 (256--681) \\
Pixel spacing (mm) & 0.69 (0.54--0.90) & 0.75 (0.62--0.98) \\
Reconstruction diameter (mm) & 352 (275--456) & 382 (317--500) \\
Mean tube current (mA) & 386 (107--495) & 243 (84--467) \\
\bottomrule
\multicolumn{3}{@{}p{0.92\linewidth}@{}}{\footnotesize Note.---Data are numbers of series or
medians with ranges in parentheses. All series are helical acquisitions reconstructed as 512
$\times$ 512 axial images with 12 bits stored. Manufacturer and scanner model were removed
during de-identification.}
\end{tabular}
\end{table*}

\begin{figure*}[t]
\centering
\includegraphics[width=0.62\linewidth]{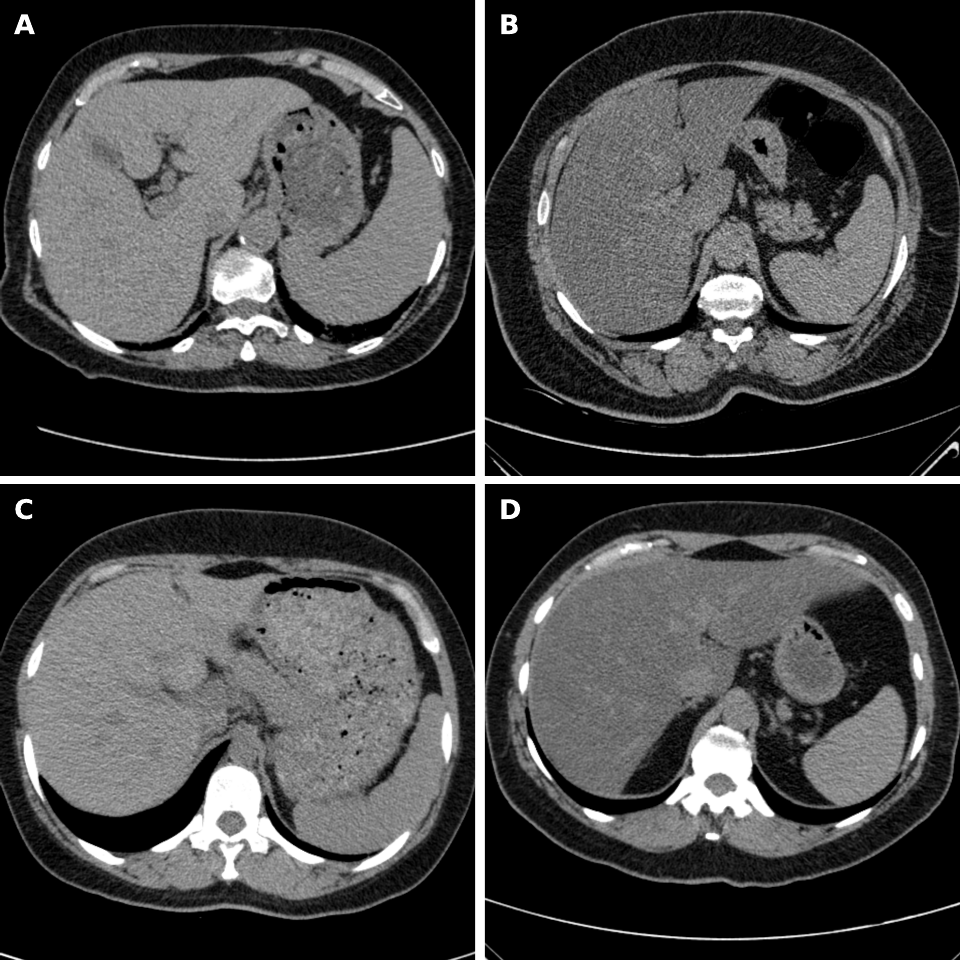}
\caption{Example unenhanced axial CT images from the training set at the level of the liver
and the spleen (window width, 350 HU; window level, 40 HU). (A) Chest CT in a patient labeled as
not having steatosis (patient 186; automated mean attenuation, 57 HU in the liver and 48 HU in
the spleen). (B) Chest CT in a patient labeled as having steatosis (patient 57; 21 HU and 52 HU);
the liver is less attenuating than the spleen and than the intrahepatic vessels. (C) Abdominal CT
in a patient labeled as not having steatosis (patient 94; 58 HU and 48 HU). (D) Abdominal CT in
a patient labeled as having steatosis (patient 6; 21 HU and 54 HU).}
\label{fig:examples}
\end{figure*}

\subsection*{Training and Test Sets and Competition Results}

The training set comprises 226 patients (114 with chest CT and 112 with abdominal CT), with 243
studies, 301 series, and 85\,464 images. The test set comprises 38 patients, all with chest CT,
with 42 studies, 43 series, and 14\,419 images (Fig~\ref{fig:flow},
Table~\ref{tab:composition}). The distribution of the labels in the test set is withheld. The
competition had 29 entrants and three teams, who made 25 submissions; the best final log loss
was 0.215. The competition remains open for late submissions.

\begin{figure*}[t]
\centering
\includegraphics[width=0.8\linewidth]{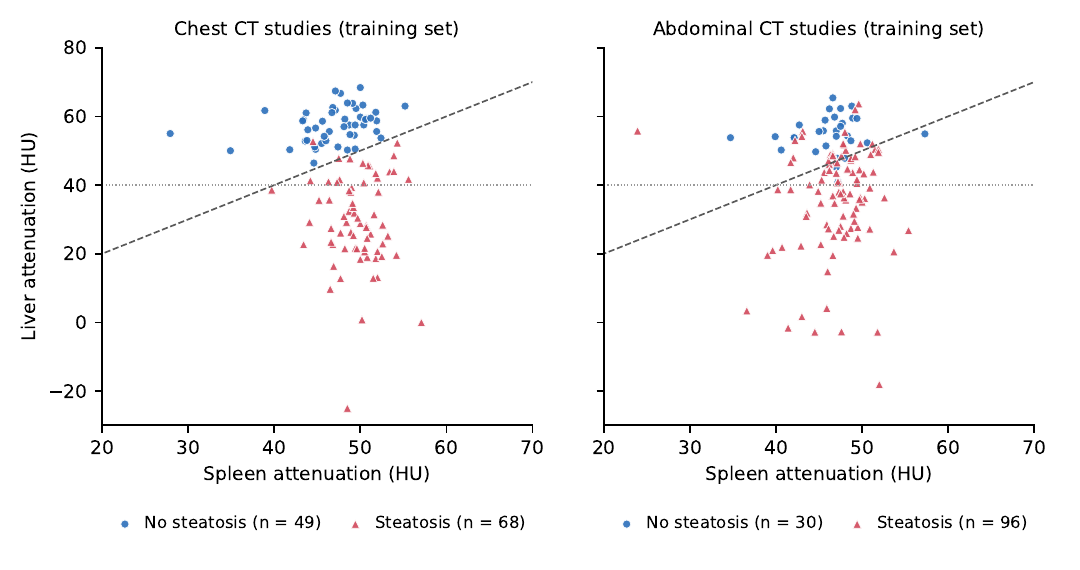}
\caption{Automated mean attenuation of the liver and the spleen in the 243 studies of the
training set, according to the patient-level label and the type of examination. Each point is
one study. The dashed line indicates equal attenuation of the two organs; points below it
fulfill the criterion of liver attenuation lower than spleen attenuation. The dotted line marks
a liver attenuation of 40 HU. Measurements were obtained with an automated segmentation model
for this report and were not used to assign the labels.}
\label{fig:scatter}
\end{figure*}

\subsection*{Image Files and Technical Characteristics}

All 99\,883 files are CT Image Storage objects, stored as uncompressed 512 $\times$ 512 axial
images with attenuation values in Hounsfield units after application of the rescale intercept
($-1024$) and slope (1). Every file holds the same 63 attributes, none of which is private, and
no file contains patient information: the attributes for dates and times, birth date, sex, age,
manufacturer, and referring physician are empty, and the attributes for patient name and
patient identification hold only the sequential case number. Every folder holds a single series
with unique section positions, and every case folder matches a case number in one of the two
comma-separated value files. In two studies (cases 59 and 267), the study folder holds one
series described as ``Chest'' and one described as ``Abdomen''. The technical characteristics of
the series are presented in Table~\ref{tab:technical}. Most chest series have a section
thickness of 1 mm (162 of 165), whereas abdominal series are mostly 3 mm (114 of 179) or 1 mm
(56 of 179), and all but one series were acquired at 120 kVp. Example images are shown in
Figure~\ref{fig:examples}.

\subsection*{Agreement Between Automated Measurements and Labels}

A segmentation of the liver, spleen, and aorta was obtained in all 285 studies; a chest series
was analyzed in 159 studies and an abdominal series in 126. Aortic attenuation was consistent
with unenhanced acquisitions (median, 41 HU; range, 24--93 HU; one study above 80 HU). The
liver mask reached the lower edge of the scan volume in 124 of the 159 chest studies (78\%) and
the spleen mask in 36 (23\%), although the volume of liver included was similar to that of
abdominal studies (median segmented volume, 1636 mL vs 1689 mL).

In the 243 studies of the training set, the median automated liver attenuation was 57 HU (IQR,
53--60 HU) in studies of patients labeled as not having steatosis and 36 HU (IQR, 25--44 HU) in
studies of patients labeled as having steatosis, and the median liver-minus-spleen difference
was 9 HU (IQR, 7--14 HU) and $-13$ HU (IQR, $-23$ to $-4$ HU), respectively
(Fig~\ref{fig:scatter}). At the study level, the automated criterion agreed with the label in
218 of 243 studies (90\%), with higher agreement for chest studies (115 of 117, 98\%) than for
abdominal studies (103 of 126, 82\%); the area under the receiver operating characteristic curve
of the liver-minus-spleen difference was 0.96. At the patient level, the automated criterion
agreed with the label in 206 of 226 patients (91\%; Cohen $\kappa$ = 0.81). Of the 20 discordant
patients, 16 were labeled as having steatosis although the automated liver attenuation was equal
to or higher than that of the spleen, and four were labeled as not having steatosis with a liver
attenuation 0.2--6.3 HU lower than that of the spleen. Among the 164 studies of patients labeled
as having steatosis, 102 (62\%) had an automated liver attenuation below 40 HU. The 15 training
patients with more than one study are all labeled as having steatosis; in these patients, the
liver-minus-spleen difference varied between studies by a median of 8 HU (range, 1--68 HU).

\section*{Discussion}

We curated a public dataset of 285 unenhanced chest and abdominal CT studies from 264 patients,
with patient-level labels for hepatic steatosis. To our knowledge, it is the first openly
downloadable CT dataset with radiologist-assigned steatosis labels that is centered on chest CT,
an examination in which the liver is only partly depicted and is rarely the focus of the
interpretation.

The design of the dataset reflects the intended application. A model that flags steatosis on
chest CT has to work with an incomplete liver and, often, with a small portion of the spleen; in
this dataset, the liver was truncated in about three-quarters of the chest studies. Abdominal CT
studies, in which the whole liver and spleen are visible, were included in the training set
only, so that they can be used as an additional source of supervision, whereas performance is
measured on chest CT alone. This arrangement makes the dataset suitable for studying the
transfer of models between examination types, in addition to the classification task itself.

The labels are provided at the patient level, without ROIs or segmentations. This makes the task
a weakly supervised one: a model receives a whole CT volume, in which the liver occupies a small
fraction of the voxels, and a single binary label. Possible approaches include end-to-end
three-dimensional classification, classification of selected sections, and pipelines that first
segment the liver and the spleen with publicly available tools \citep{wasserthal2023} and then
classify attenuation statistics. The last approach also allows the dataset to be used as an
independent test set for automated liver fat quantification methods developed elsewhere
\citep{graffy2019,jirapatnakul2020,modanwal2022}.

The technical validation supports the consistency of the labels: a fully automated version of
the annotation criterion reproduced the label in 91\% of the training patients. We did not
adjudicate the discordant cases. Possible explanations include the difference between three
manually placed ROIs and a whole-organ average, heterogeneous fat deposition, and segmentation
errors; users of the dataset should nevertheless expect some label noise, particularly among
abdominal studies, in which agreement was lower. The labels also include mild changes: more than
one-third of the studies of patients labeled as having steatosis had a liver attenuation of 40
HU or higher, whereas an attenuation below 40 HU is the threshold commonly used to define
moderate to severe steatosis \citep{chen2017,medeiros2020,yardeni2026}. The small number of participants in the competition means
that its results should not be taken as an estimate of the performance that can be achieved on
this task.

There are several limitations of this dataset. First, it comes from a single institution, and
the scanner manufacturer and model, the patient age and sex, and the examination dates are not
available, so that the composition of the sample cannot be described and subgroup analyses are
not possible. Second, the labels were assigned by a single reader who was a resident at the
time, without a second reading or a measure of interobserver agreement, and only the binary
result was recorded; the individual ROI measurements are not available. Third, the reference
standard is CT attenuation relative to the spleen, not histologic examination or MRI proton
density fat fraction; CT criteria have low sensitivity and are inaccurate for mild steatosis
\citep{pickhardt2012,lee2014}, and the liver-to-spleen comparison is affected by
conditions that change the attenuation of either organ, such as iron overload or amiodarone
deposition. Fourth, the sample is not consecutive and is enriched for steatosis (65\% of training
patients), so that prevalence-dependent metrics do not transfer to clinical populations. Fifth,
the label is assigned to the patient, yet 18 patients have more than one study, and liver fat
content can change between examinations. Sixth, the test set is small (38 patients), which
results in wide confidence intervals for any performance estimate. Finally, examinations with
large or diffuse liver lesions were excluded, and their number was not recorded, so models
developed with this dataset have not been exposed to such cases.

In conclusion, we created and made publicly available a dataset of 285 unenhanced chest and
abdominal CT studies from 264 patients with patient-level labels for hepatic steatosis, released
it through a machine learning competition, and described its composition; an automated version
of the annotation criterion agreed with the labels in 91\% of the training patients. The dataset
is freely available at \url{\kaggleurl}.

\section*{Acknowledgments}

We thank Ernandez Rodrigues, administrator of the picture archiving and communication system,
and Jefferson Silva, information technology manager, for their support in data extraction. This work was presented in part as a poster at the
European Society of Medical Imaging Informatics (EuSoMII) Annual Meeting 2022 (``The UNIFESP CT
Chest Fatty Liver Dataset and Competition'').

\section*{Funding and Disclosures}

The authors received no specific funding for this work and declare no conflicts of interest.

\end{document}